\documentclass{webofc}
\usepackage[varg]{txfonts} 

\usepackage{amssymb}
\usepackage{amsmath}

\usepackage{graphics}
\usepackage{epsfig}

\begin{document} %



\title{Bound on a flux of ultra-high energy neutrinos in a scenario with
extra dimensions}

\author{%
\firstname{Mikhail} \lastname{Astashenkov}\inst{1}
\fnsep\thanks{\email{mixa.astash@yandex.ru}} \and
\firstname{Alexander} \lastname{Kisselev}\inst{2}\fnsep
\thanks{\email{alexandre.kisselev@ihep.ru}}
}

\institute{Department of Physics, Lomonosov Moscow State University,
119991 Moscow, Russia \and A.A.~Logunov Institute for High Energy
Physics, NRC ``Kurchatov Institute'', 142281 Protvino, Russia}

\abstract{%
Assuming that a single-flavor diffuse neutrino flux $dN_\nu/dE_\nu$
is equal to $k E_\nu^{-2}$ in the energy range $10^{17}$ eV -- $2.5
\times 10^{19}$ eV, an upper bound on $k$ is calculated in the ADD
model as a function of the number of extra dimensions $n$ and
gravity scale $M_D$. An expected number of neutrino induced events
at the Surface Detector array of the Pierre Auger Observatory is
estimated.}

\maketitle



\section{Introduction}
\label{sec:intr}

High energy cosmic neutrinos may help us \\
-- to discover cosmic rays (CRs) point sources; \\
-- to define their position, in particular, to constrain a
  position of the gravity wave (GW) sources; \\
-- to understand a mechanisms of CR acceleration; \\
-- to define an energy boundary between galactic and
extragalactic parts of CR spectrum; \\
-- to give information on a nature of CR composition;\\
-- to measure a cosmic neutrino flux, flavor ratio and
  high energy neutrino-nucleon cross section.

The first observation of high-energy astrophysical neutrinos was
done by the IceCube neutrino detector \cite{IceCube:2014}. The
single-flavor diffuse neutrino flux was measured in the energy
region $25 \mathrm{\ TeV} < E_\nu < 1.4 \mathrm{\ PeV}$ to be
\cite{IceCube:2015}
\begin{equation}\label{IceCube_flux}
\frac{dN}{dE_\nu} = 2.06 \times 10^{-18} \left( \frac{E_0}{E_\nu}
\right)^{\!\!\gamma} \mathrm{GeV^{-1} cm^{-2} s^{-1} sr^{-1} } ,
\end{equation}
where $E_0 = 10^5$ GeV, $\gamma = 2.46$. Later on, it was found that
the neutrino-nucleon cross section agrees with SM predictions in the
range 6.3 TeV -- 980 TeV \cite{IceCube:2017}.

Ultra-high energy (UHE) neutrinos (with energies above $10^{17}$ eV)
are of particular interest. They may probe a new physics if the
latter gives us a significant enhancement of neutrino-nucleon cross
sections. To detect UHE neutrinos, powerful CR detectors, such as
the Pierre Auger Observatory (PAO) \cite{PAO} or Telescope Array
\cite{TA}, are needed. Recently, the PAO Collaboration reported on
searches both for downward-going (DG) and Earth-skimming (ES)
neutrinos \cite{Auger:2015}. The DG air showers
\cite{Berezinsky:1969} are initiated by cosmic neutrinos moving with
large zenith angle which interact in the atmosphere near the Surface
Detector (SD) array of the PAO. The ES air showers
\cite{Bertou:2002} are induced by tau neutrinos coming at small
negative angles with respect to the horizon which interact in the
Earth producing tau leptons. In their turn, the tau leptons escape
the Earth and initiate showers close to the SD. The zenith angles of
$60^\circ - 75^\circ$ and $75^\circ - 90^\circ$ for the DG air
showers, and zenith angles of $90^\circ - 95^\circ$ for the ES air
showers were analyzed. The data were collected for a period which is
equivalent of 6.4 years of a complete PAO SD working continuously
\cite{Auger:2015}.

No neutrino candidates were found. Assuming the diffuse flux of UHE
neutrinos to be
\begin{equation}\label{flux_en_dependence}
\frac{dN}{dE_\nu} = k \,E_\nu^{-2}
\end{equation}
in the energy range $1.0\times10^{17}$ eV -- $2.5\times10^{19}$ eV,
the single-flavor  upper limit to the diffuse flux of UHE neutrinos
was obtained by the PAO Collaboration \cite{Auger:2015}
\begin{equation}\label{Auger_bound}
k < 6.4 \times 10^{-9} \mathrm{\ GeV \ cm^{-2} \ s^{-1} \ sr^{-1}} .
\end{equation}
This bound is approximately four times less than the benchmark
Waxman-Bachall bound on cosmic neutrino production in optically thin
sources \cite{WB:2001}. Note that the IceCube flux
\eqref{IceCube_flux}, if extrapolated to 1 EeV, would give
\begin{equation}\label{IceCube_flux_ext}
k = 0.3 \times 10^{-9} \mathrm{\ GeV \ cm^{-2} \ s^{-1} \ sr^{-1}} .
\end{equation}
Recently, from the nonobservation of neutrino candidates from the GW
sources \cite{GW} the following upper limit was derived
\cite{Auger_GW}, \cite{Kotera:2016}
\begin{equation}\label{GW_flux}
E_\nu^2 \frac{dN}{dE_\nu} = (1.5 - 6.9) \times 10^{-8} \mathrm{\ GeV
\ cm^{-2} \ s^{-1} \ sr^{-1}} .
\end{equation}

The bound \eqref{Auger_bound} has was obtained under assumption that
neutrino-nucleon collisions in the atmosphere are described by the
CC and NC interactions. The goal of the present paper is to
calculate the single-flavor upper bound on the diffuse flux of UHE
cosmic neutrinos in the ADD model \cite{Arkani-Hamed:98} as a
function of the number of extra dimensions $n$ and $D$-dimensional
Planck scale $M_D$ ($D=4+n$).


\section{Neutrino-nucleon cross sections}
\label{sec:nu_N}

Consider energy region $E_\nu > 10^{17}$ eV. At UHEs the neutrino
interacts essentially with the the partons (quarks, antiquarks and
gluons) inside the nucleon. If the impact parameter of the incoming
neutrino $b$ is much larger than a $D$-dimensional Schwarzschild
radius $R_S$ \cite{Myers:86},
\begin{equation}\label{R_S}
R_S(\hat{s}) = \frac{1}{\sqrt{\pi}} \frac{1}{M_D} \left[ \frac{8
\Gamma \left( \frac{n+3}{2}\right) }{n+2} \frac{\sqrt{\hat{s}}}{M_D}
\right]^{\frac{1}{n+1}} ,
\end{equation}
the eikonal approximation for the scattering amplitude
\cite{Cheng:69} is valid. Here $\sqrt{\hat{s}}$ is an invariant
energy of the partonic subprocess. Then we are in so-called
transplanckian regime \cite{Giudice:02} which corresponds to the
conditions
\begin{equation}\label{tarns_Pl_regime}
\sqrt{\hat{s}} \, \gg M_D \;, \quad \theta \sim \left[ \frac{R_S
\!\left(\hat{s} \right)}{b} \right]^{n+1} \ll 1 ,
\end{equation}
were $\theta$ is a scattering angle.

In the eikonal approximation the leading part of the scattering
amplitude is obtained by summation of all ladder diagrams with
graviton exchange in the $t$-channel. The tree-level exchanges of
the $D$-dimensional graviton gives the eikonal formula
\begin{equation}\label{A_eik}
A_{\mathrm{eik}}(s,t) = -2is \!\int \!\!d^2b \,e^{iq b} \left[ e^{i
\chi(b)} - 1 \right] ,
\end{equation}
with the eikonal phase \cite{Giudice:02} - \cite{Emparan:02}
\begin{equation}\label{eik_final}
\chi(b) = \left( \frac{b_c}{b} \right)^{\!n} \;, \quad \text{where \
} b_c = \left[ \frac{(4\pi)^{n/2-1} \hat{s} \Gamma(n/2)}{2M_D^{n+2}}
\right]^{1/n} .
\end{equation}

If the impact parameter of the incoming neutrino is less than $R_S$,
the neutrino and the parton form a black hole. In such a case, the
cross section can be estimated as \cite{Feng:02} -
\cite{Giddings:02}
\begin{equation}\label{CS_bh}
\sigma_{\nu N \rightarrow \mathrm{BH}}(s) = \pi \sum_{i}
\!\!\int\limits_{(M_{\mathrm{bh}}^{\mathrm{min}})^2/s}^1 \!\!\!\!dx
f_i (x, \hat{s}) \,R_S^2(\hat{s}) \;,
\end{equation}
where $f_i (x, \hat{s})$ is the parton distribution function (PDF)
of the parton $i$ with the momentum fraction $x$ inside the nucleon
($i=q, \bar{q}, g$), and $\hat{s} = x s$. We use the CT14 set of the
PDFs \cite{CT14}. For chosen value of $n$ we take $M_D$ to be equal
to a lower limit on $M_D$ obtained by CMS Collaboration (see fig.~11
in \cite{CM:limits_MD}). For given values of $n$ and $M_D$, we fix
$M_{\mathrm{bh}}^{\mathrm{min}}$ to be equal to a corresponding
lower limit on $M_{\mathrm{bh}}^{\mathrm{min}}$ from
ref.~\cite{CM:limits_MBH}.

\begin{figure}[htb]
\begin{center}
\includegraphics[width=6cm,clip]{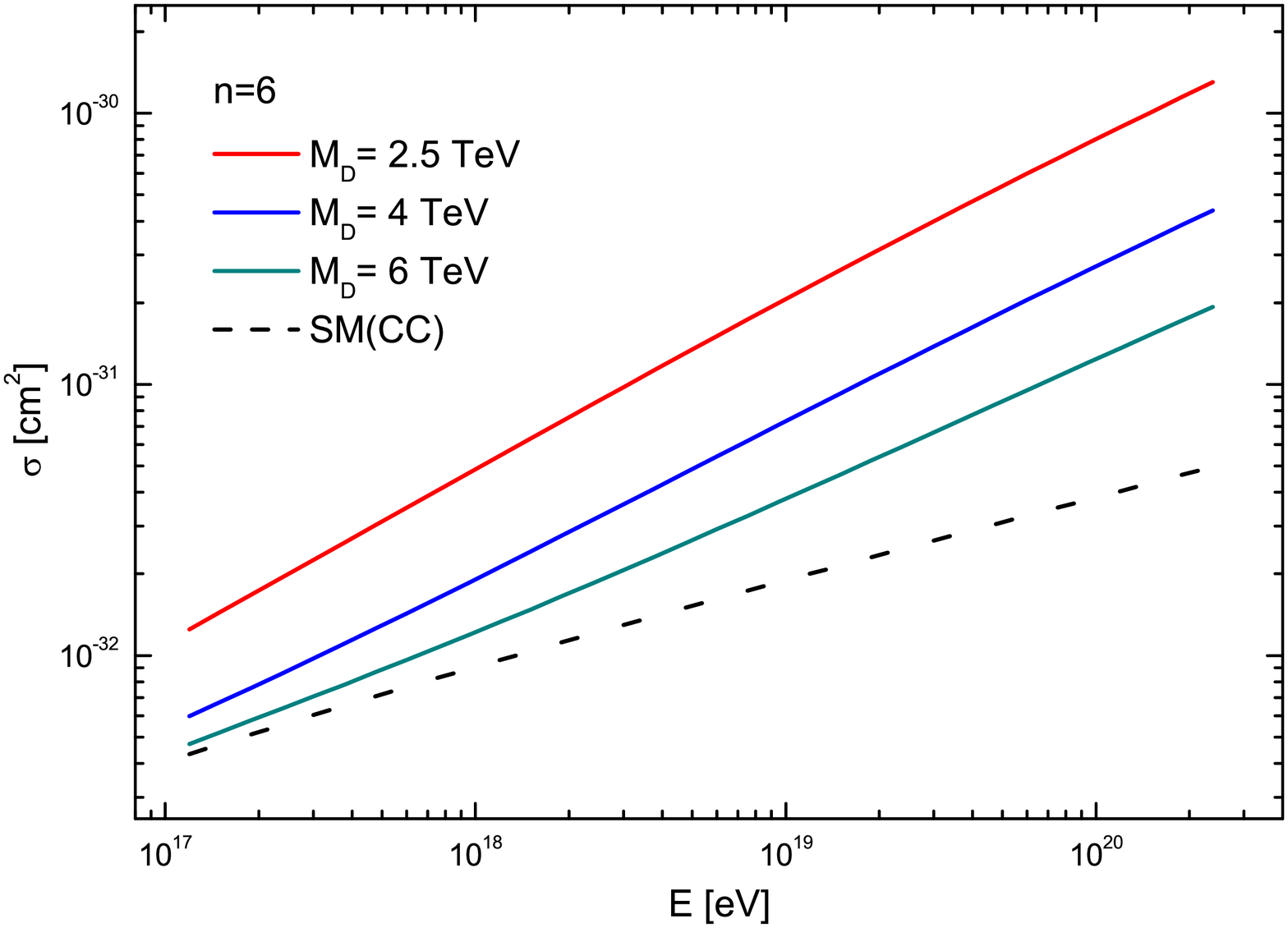} \hspace{.5cm}
\includegraphics[width=6cm,clip]{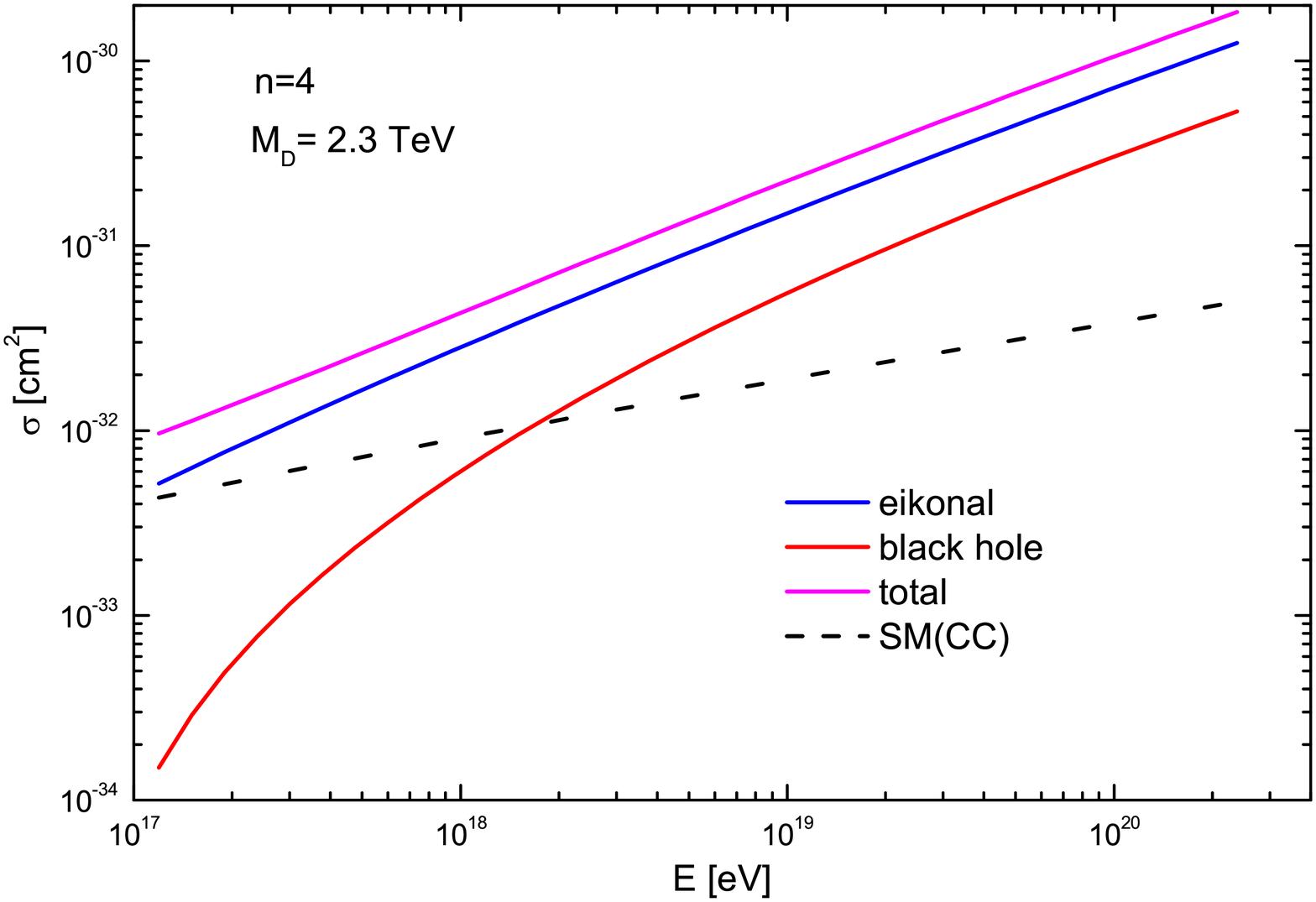}
\caption{Left panel: the neutrino-nucleon total cross sections for
$n=6$ and $M_D = 2.5$ TeV, 4.0 TeV, 6.0 TeV for the neutrino
energies above $10^{17}$ eV. Right panel: the eikonal and black hole
contributions to the total cross section for $n=4$.}
\label{fig:cross_sections}
\end{center}
\end{figure}

As for the SM cross section for the UHE neutrino-nucleon scattering
off the nucleon, we adopt a formula from \cite{Sarkar:2008}, which
was used by the PAO Collaboration to obtain the upper limit on the
neutrino flux \eqref{Auger_bound}. The total cross sections as a
function of $M_D$ and fixed $n$ are shown in
figs.~\ref{fig:cross_sections}. The eikonal and black hole
contributions to the total cross section are also shown for $n=4$,
$M_D=2.3$ TeV.
%

In fig.~{\ref{fig:cross_sections_IceCube} the neutrino-nucleon total
cross section for the energy range $10^{13}$ eV -- $10^{17}$ eV is
shown. As one can see, for the sensitivity region of the detector
IceCube ($E_\nu \lesssim 2$ PeV) effects from the extra dimensions
are negligible, in accordance with the cross section measurements by
the IceCube Collaboration \cite{IceCube:2017}. But they become
important at $E_\nu > 10^{16}$ eV.
%
\begin{figure}[htb]
\begin{center}
\includegraphics[width=6cm,clip]{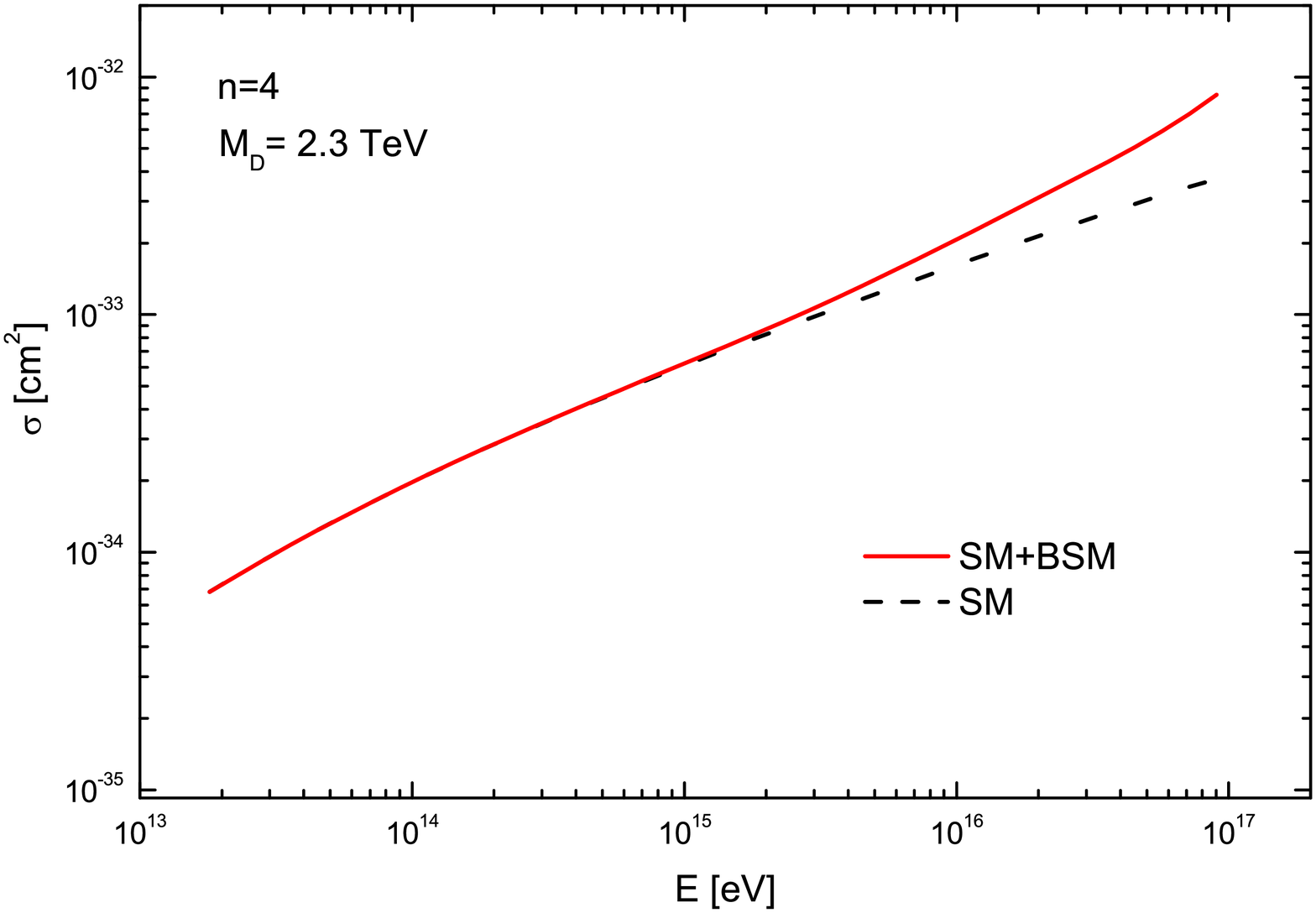}
\caption{The neutrino-nucleon total cross section for $n=4$, $M_D =
2.3$ TeV for the neutrino energies below $10^{17}$ eV in the ADD
model (solid line). The SM prediction is shown by the dashed line.}
\label{fig:cross_sections_IceCube}
\end{center}
\end{figure}

The calculation of the cross sections is not an end of the story. It
enables us to calculate exposures both for DG and ES neutrino events
at the SD array of the PAO in the ADD model and thus to put an upper
limit on the diffuse single-flavor flux of UHE neutrinos. It will be
done in the next section.


\section{Exposures and bounds on diffuse neutrino flux in the ADD model}
\label{sec:limits}

The exposure of the DG neutrino events increases with the rise of
the neutrino-nucleon cross section, that results in its dependence
on the ``new physics'' cross section $\sigma_{\mathrm{NP}}$
\cite{Kisselev:2016}
\begin{equation}\label{DG:BSM_vs_SM}
\mathcal{E}_{\mathrm{BSM}}^{\mathrm{DG}} (E_\nu) =
\mathcal{E}_{\mathrm{SM}}^{\mathrm{DG}} (E_\nu)\,
\frac{\sigma_{\mathrm{SM}}^{\mathrm{eff}}(E_\nu) +
\sigma_{\mathrm{NP}}(E_\nu)}{\sigma_{\mathrm{SM}}^{\mathrm{eff}}(E_\nu)}
\;,
\end{equation}
where $\mathcal{E}_{\mathrm{BSM}}^{\mathrm{DG}}$
($\mathcal{E}_{\mathrm{SM}}^{\mathrm{DG}}$) is the exposure of the
SD of the PAO with (without) account of the new interaction. The
effective SM cross section $\sigma_{\mathrm{SM}}^{\mathrm{eff}}$
takes into account relative mass apertures for charged current (CC)
and neutral current (NC) interactions of the DG neutrinos at the PAO
(see \cite{Kisselev:2016} for details). In contrast to
$\mathcal{E}_{\mathrm{BSM}}^{\mathrm{DG}}$, the exposure of the ES
neutrino events decreases with the rise of the neutrino total cross
section as \cite{Kisselev:2016}
\begin{equation}\label{ES:BSM_vs_SM}
\mathcal{E}_{\mathrm{BSM}}^{\mathrm{ES}}(E_\nu) =
\mathcal{E}_{\mathrm{SM}}^{\mathrm{ES}}(E_\nu) \,
\frac{\sigma_{\mathrm{CC}}^2(E_\nu)}{[\sigma_{\mathrm{CC}}(E_\nu) +
\sigma_{\mathrm{NP}}(E_\nu)]^2} \;.
\end{equation}
%
\begin{figure}[htb]
\centering
\includegraphics[width=6cm,clip]{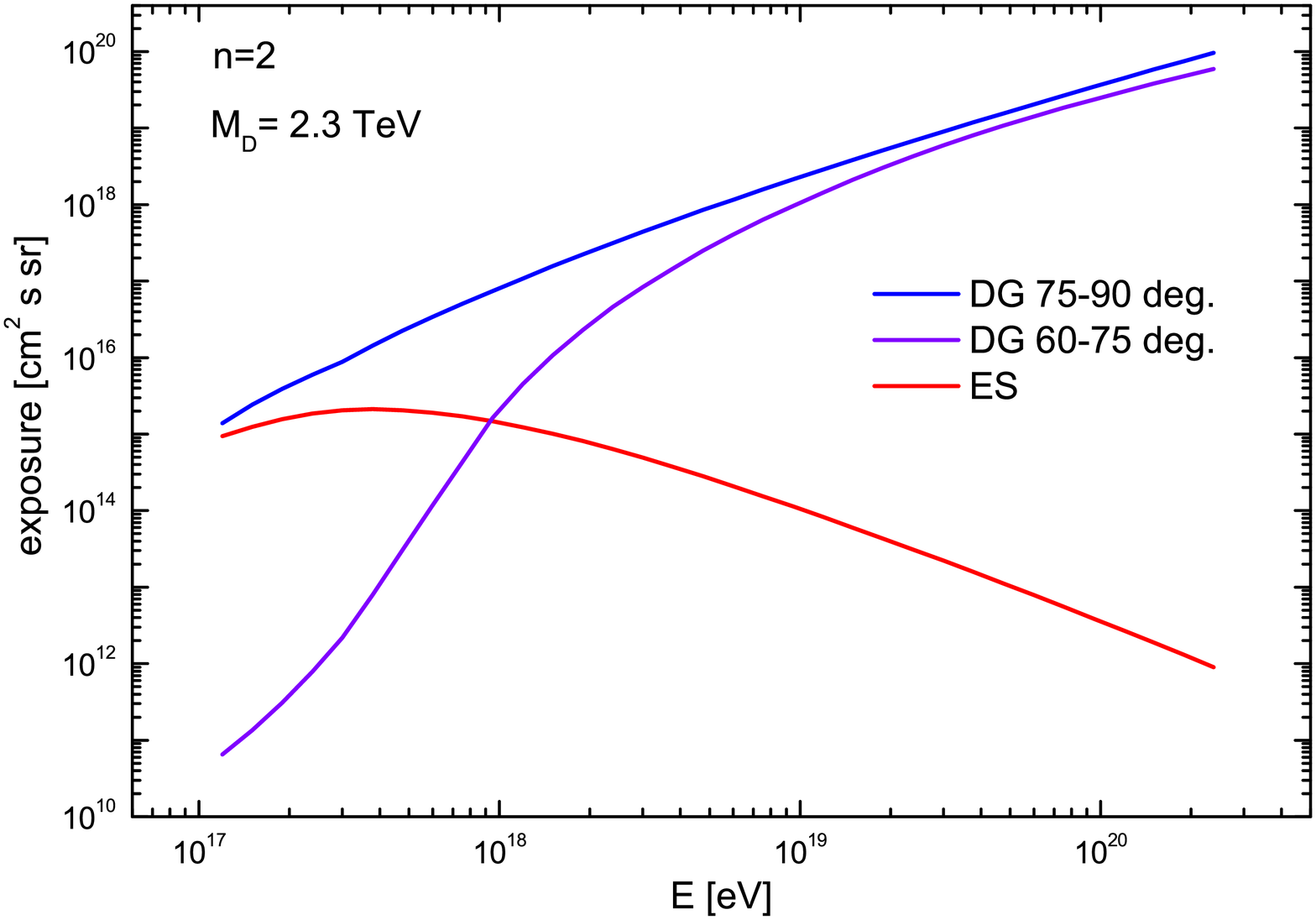} \hspace{5mm}
\includegraphics[width=6cm,clip]{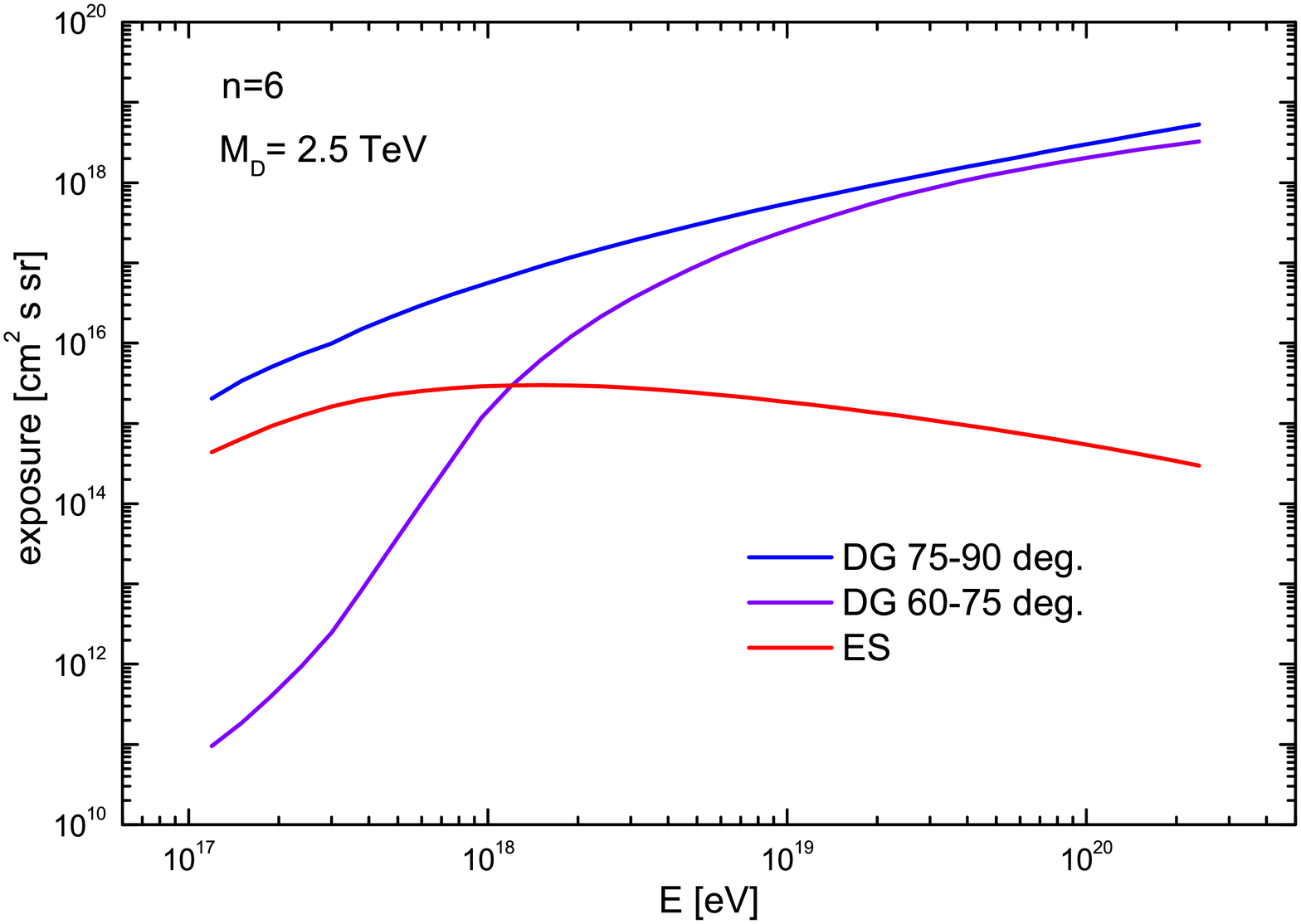}
\caption{Left panel: the expected exposures of the SD array of the
PAO for the DG and ES neutrinos in the ADD model for the period of
6.4 years. Right panel: the same as on the left panel, but for
$n=6$.} \label{fig:exposures}
\end{figure}

The formulas \eqref{DG:BSM_vs_SM} and \eqref{ES:BSM_vs_SM} allowed
us to calculate exposures of the SD of the PAO for the period 1
January 2004 -- 20 June 2013 expected in the ADD model. This period
is equivalent of 6.4 years of a complete SD array working
continuously. The PAO data on the exposures for the SM neutrino
interactions in the region from $\log(E_\nu/\mathrm{eV}) = 17$ to
20.5 were used ~\cite{Auger:2015}. The results of our calculations
are presented in figs.~\ref{fig:exposures}.

It was assumed that the astrophysical flux arrives isotropically
from all directions, and neutrino flavor composition is $\nu_e :
\nu_\mu :\nu_\tau  = 1 : 1 : 1$. Following Pierre Auger
Collaboration, we also adopted that the flux is described by a power
law of the form \eqref{flux_en_dependence}. As one can see in
fig.~\ref{fig:cross_sections}, in the ADD model the cross sections
rise more rapidly with the neutrino energy than the SM cross
sections. As a result, the exposure for the DG events,
$\mathcal{E}_{\mathrm{BSM}}^{\mathrm{DG}}$ \eqref{DG:BSM_vs_SM},
rises, while the exposure for the ES events,
$\mathcal{E}_{\mathrm{BSM}}^{\mathrm{ES}}$ \eqref{ES:BSM_vs_SM},
decreases as $E_\nu$ grows (see fig.~\ref{fig:exposures}).

The upper limit on the value of $k$ is defined as \cite{Auger:2015}
\begin{equation}\label{k_int}
k = \frac{N_{\mathrm{up}}}{\int \!E_\nu^{-2}
\mathcal{E}_{\mathrm{tot}}(E_\nu) d E_\nu} \;,
\end{equation}
where $N_{\mathrm{up}} = 2.39$ is an actual value of the upper limit
on the signal events, assuming a number of expected background
events to be zero. The results of our calculations of the upper
bound on the neutrino flux normalization $k$ are shown in
figs.~\ref{fig:k_MD_fixed_n}, \ref{fig:k_n fixed_MD}, in which the
PAO upper bound on $k$ is also shown.
%
\begin{figure}[htb]
\centering
\includegraphics[width=6cm,clip]{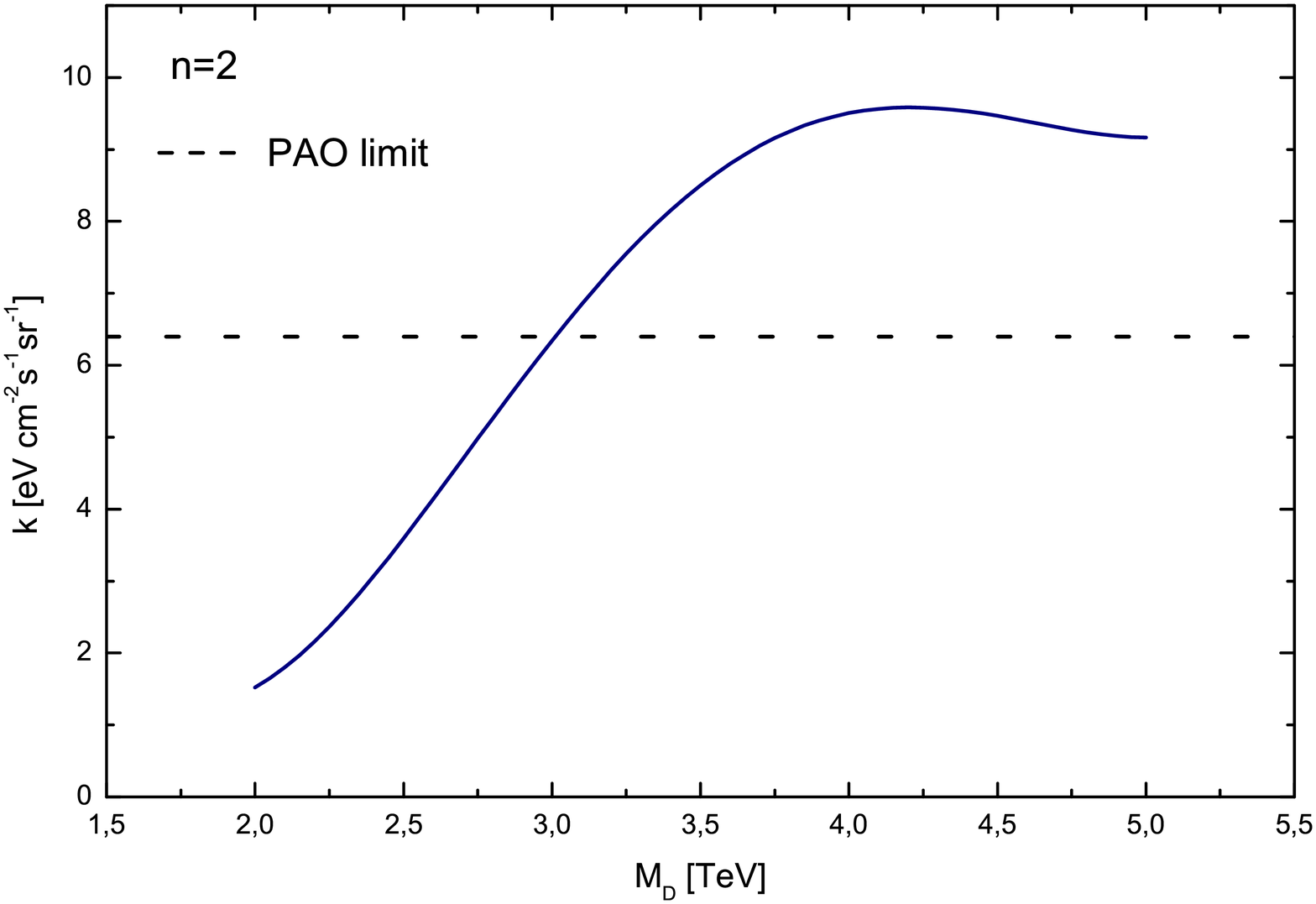} \hspace{5mm}
\includegraphics[width=6cm,clip]{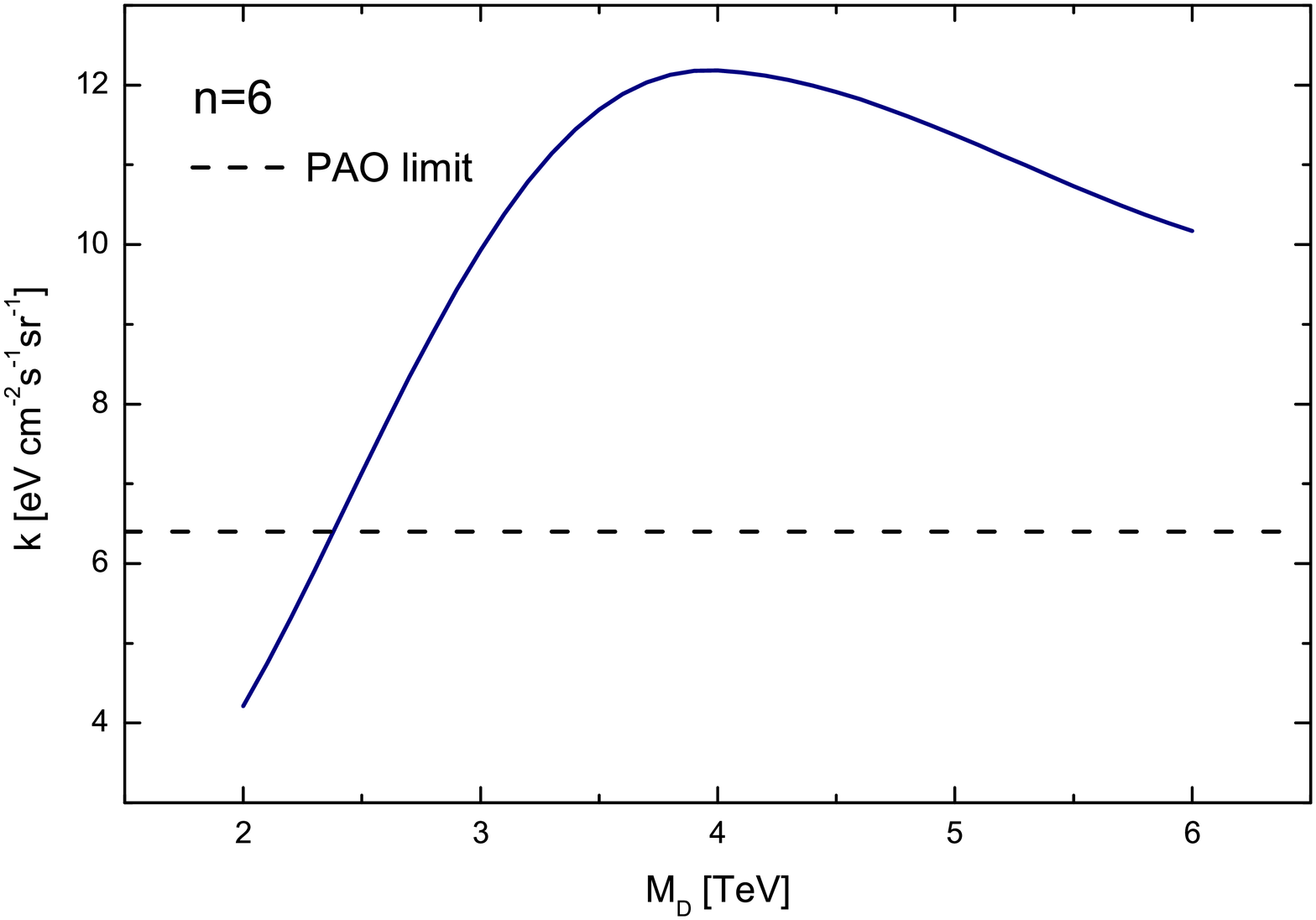}
\caption{Left panel: the upper bound on the value of $k$ as a
function of $D$-dimensional Planck scale $M_D$ for $n=2$ in the ADD
model. Right panel: the same as on the left panel, but for $n = 6$.}
\label{fig:k_MD_fixed_n}
\end{figure}
%
\begin{figure}[htb]
\centering
\includegraphics[width=6cm,clip]{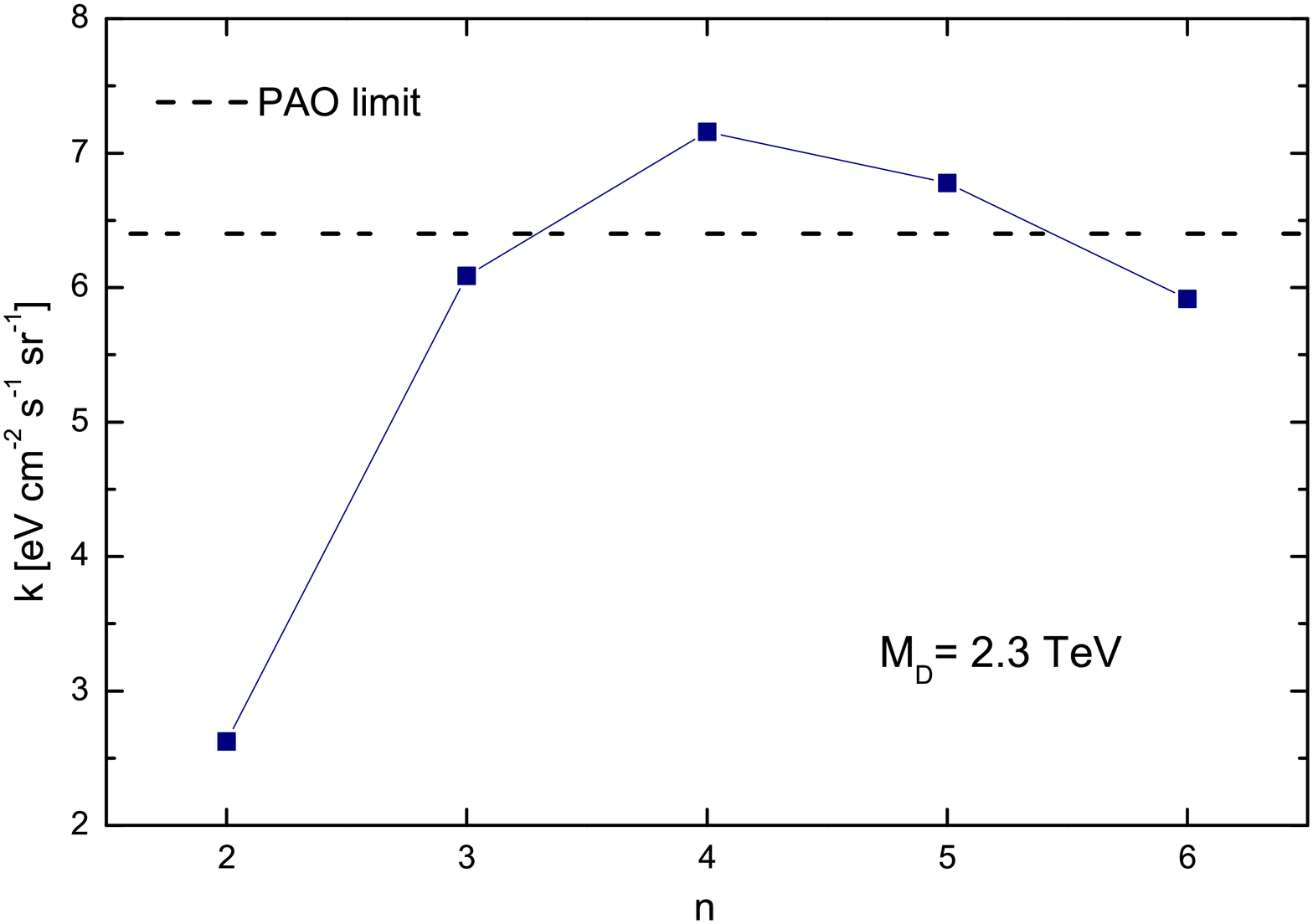} \hspace{5mm}
\includegraphics[width=6cm,clip]{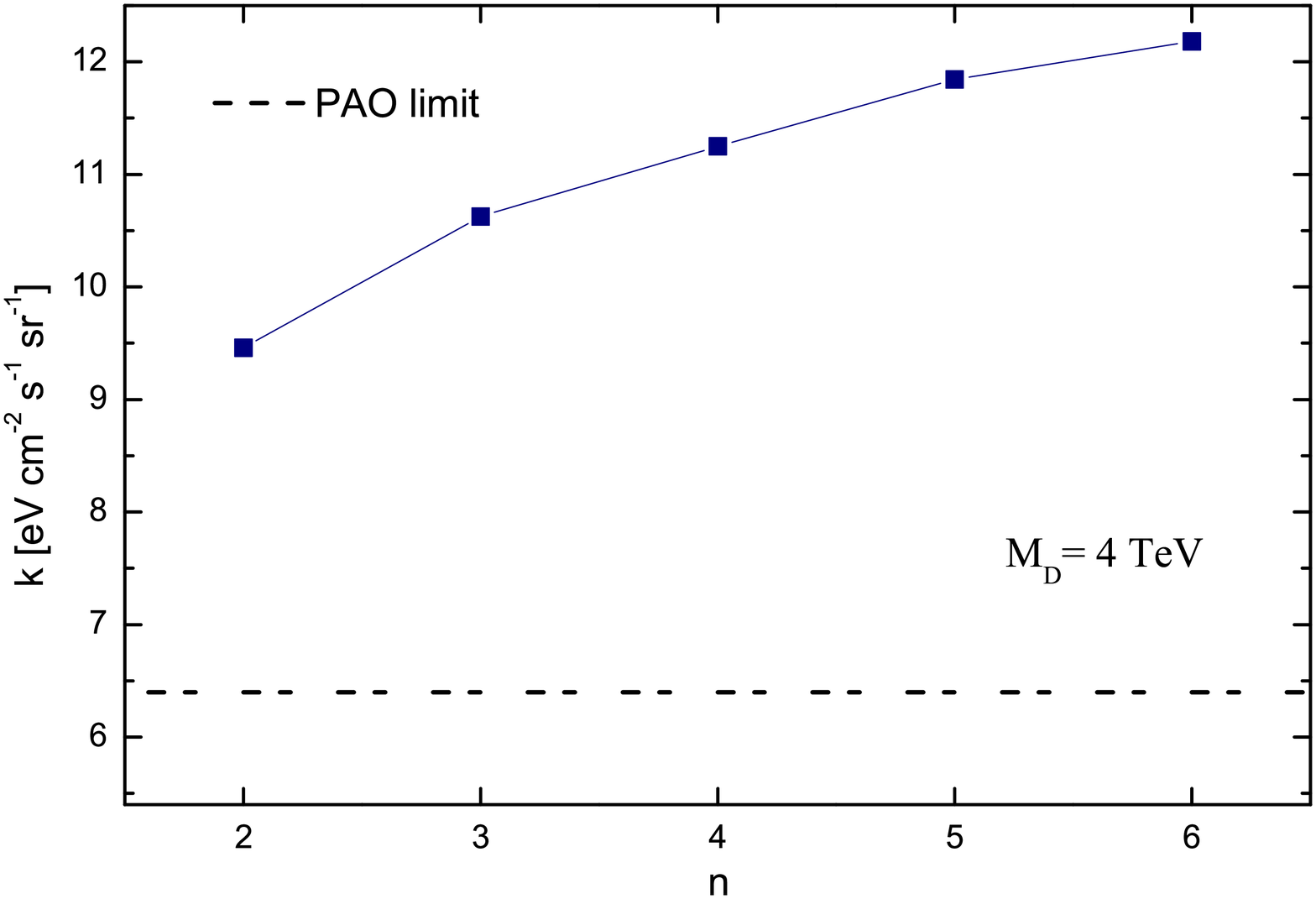}
\caption{Left panel: the upper bound on the value of $k$ as a
function of number of extra dimensions $n$ for $M_D= 2.3$ TeV. Right
panel: the same as on the left panel, but for $M_D= 4.0$ TeV.}
\label{fig:k_n fixed_MD}
\end{figure}

Finally, we have estimated the expected numbers of the neutrino
events at the SD of the PAO for the period of $2\times6.4$ years.
The calculations were done for the IceCube flux \eqref{IceCube_flux}
extrapolated to the UHE region. The results are presented in
fig.~\ref{fig:event_number_MD_fixed_n}.
%
\begin{figure}[htb]
\centering
\includegraphics[width=6cm,clip]{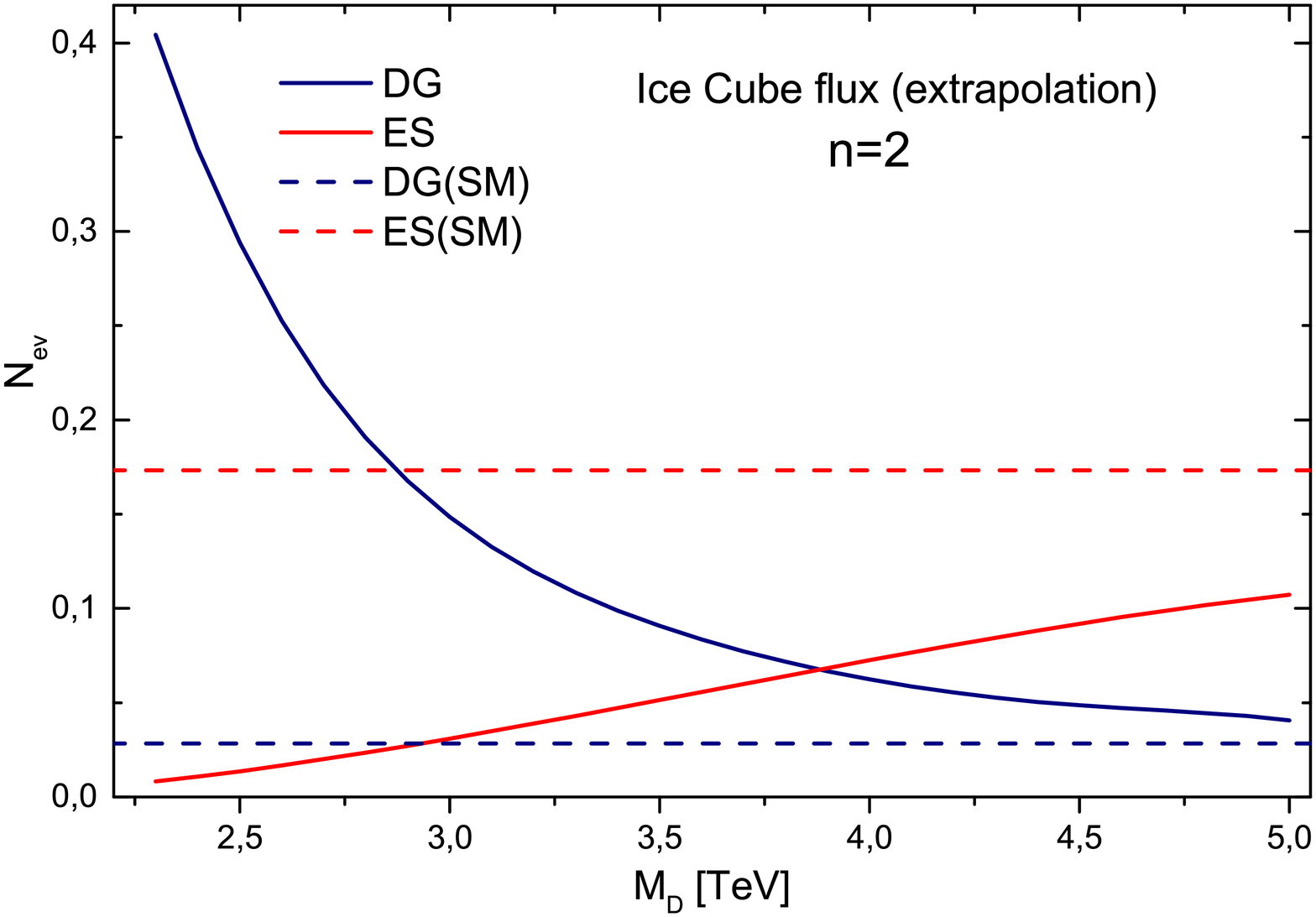} \hspace{5mm}
\includegraphics[width=6cm,clip]{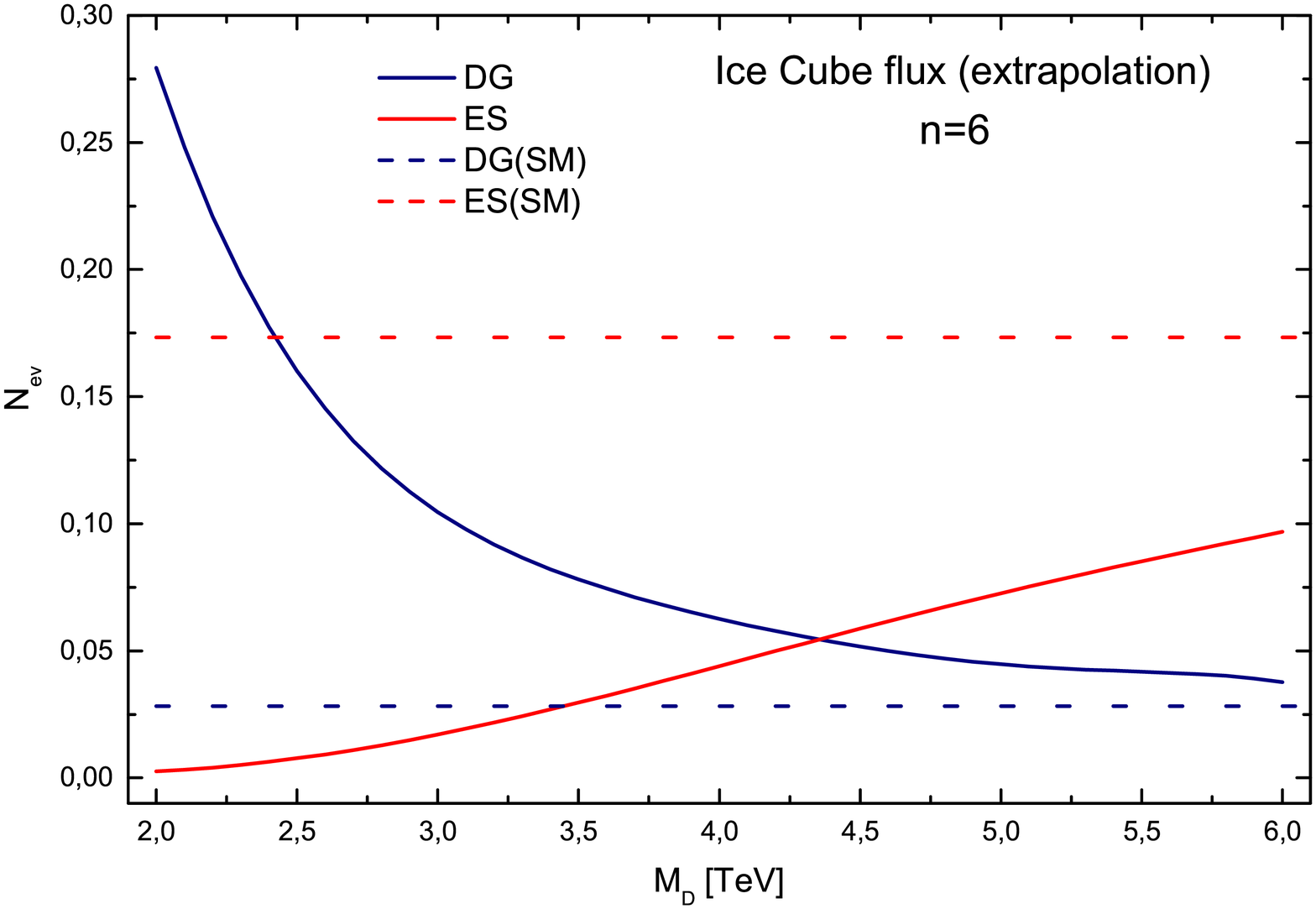}
\caption{Left panel: the expected number of events at the SD array
of the PAO for the period of $2\times6.4$ years for the IceCube
neutrino flux extrapolated to EeV energy region. Right panel: the
same as on the left panel, but for $n=6$.}
\label{fig:event_number_MD_fixed_n}
\end{figure}



\section{Conclusions}

Using the exposure of the PAO for the period equivalent of 6.4 years
of the complete PAO SD array working continuously, we have estimated
the PAO exposures for the neutrino induced events expected in the
ADD model with $n$ extra dimensions and gravity scale $M_D$. Both
downward-going and Earth-skimming UHE cosmic neutrinos were
considered. As a result, we have calculated the single-flavor upper
limit on the diffuse neutrino flux in the presence of the massive
graviton interactions in the ADD model. We assumed that the flux of
UHE neutrinos has the power form \eqref{flux_en_dependence}.

It appeared that in the ADD model the upper bound on the diffuse
neutrino flux can be more stringent that the PAO bound
\eqref{Auger_bound}, depending on two parameters of the ADD model.
As one can see in fig.~\ref{fig:k_MD_fixed_n}, it takes place for
$M_D < 3.01$ TeV (2.38 TeV), if $n=2$ (6). For $M_D = 2.3$ TeV it is
true for $n \leqslant 3$ and $n \geqslant 6$ (see the left panel of
fig.~\ref{fig:k_n fixed_MD}). However, with the increase of $M_D$
our bound becomes weaker than the PAO bound for all $n$, and it
tends to it from above as $M_D$ grows (see, for instance, the right
panels of figs.~\ref{fig:k_MD_fixed_n}, \ref{fig:k_n fixed_MD}). All
these results are explained by different dependence of the DG and ES
exposures on the neutrino-nucleon cross section (see formulas
\eqref{DG:BSM_vs_SM}, \eqref{ES:BSM_vs_SM}).

In the ADD model the expected numbers of the neutrino events
$N_{\mathrm{ev}}$ at the SD of the PAO are calculated for the period
of $2\times6.4$ years for the IceCube flux \eqref{IceCube_flux}
extrapolated to the UHE region (see
fig.~\ref{fig:event_number_MD_fixed_n}). These numbers tend to the
SM predictions, as $M_D$ grows. For the ES events, the SM value is
achieved for very large values of $M_D$.




\end{document}